\documentclass[epsfig,amsmath,amssymb,twocolumn, usenatbib,a4paper]{mn2e} 
\usepackage{graphicx}
\usepackage{dcolumn}
\usepackage{bm}
\usepackage{natbib}
\usepackage{color}
\usepackage{times}

\fontencoding{T1}

\newcommand{\be}{\begin{equation}}
\newcommand{\ee}{\end{equation}}
\newcommand{\ba}{\begin{eqnarray}}
\newcommand{\ea}{\end{eqnarray}}

\newcommand{\LCDM}{$ \Lambda $CDM}

\newcommand{\fnl}{f_{\mathrm{NL}}}

\title[Constraining non-Gaussianity with CMB lensing and LSS]{Using correlations between CMB lensing and large-scale structure to measure primordial non-Gaussianity}

\author[T.~Giannantonio \& W.~J.~Percival]{Tommaso Giannantonio$^{1,2}$\thanks{tommaso.giannantonio@usm.lmu.de} and Will J. Percival$^3$\\
$^1$Ludwig-Maximilians-Universit\"at M\"unchen, Universit\"ats-Sternwarte M\"unchen, Scheinerstr. 1, D-81679 M\"unchen, Germany\\
$^2$Excellence Cluster Universe, Boltzmannstr. 2, D-85748 Garching bei M\"unchen, Germany\\
$^{3}$Institute of Cosmology and Gravitation, University of Portsmouth, Portsmouth, PO1 3FX, UK
}

\begin{document}

\date{\today}

\pagerange{\pageref{firstpage}--\pageref{lastpage}} \pubyear{2013}

\maketitle

\label{firstpage}

\begin {abstract} 
We apply a new method to measure primordial non-Gaussianity, using the cross-correlation between galaxy surveys and the CMB lensing signal to measure galaxy bias on very large scales, where local-type primordial non-Gaussianity predicts a $k^2$ divergence. We use the CMB lensing map recently published by the Planck collaboration, and measure its external correlations with a suite of six galaxy catalogues spanning a broad redshift range. We then consistently combine correlation functions to extend the recent analysis by Giannantonio \textit{et al.} (2013), where the density-density and the density-CMB temperature correlations were used. Due to the intrinsic noise of the Planck lensing map, which affects the largest scales most severely, we find that the constraints on the galaxy bias are similar to the constraints from density-CMB temperature correlations. Including lensing constraints only improves the previous statistical measurement errors marginally, and we obtain $ \fnl = 12 \pm 21 $ (1$\sigma$) from the combined data set. However, the lensing measurements serve as an excellent test of systematic errors: we now have three methods to measure the large-scale, scale-dependent bias from a galaxy survey: auto-correlation, and cross-correlation with both CMB temperature and lensing. As the publicly available Planck lensing maps have had their largest-scale modes at multipoles $l<10$ removed, which are the most sensitive to the scale-dependent bias, we consider mock CMB lensing data covering all multipoles. We find that, while the effect of $\fnl$ indeed increases significantly on the largest scales, so do the contributions of both cosmic variance and the intrinsic lensing noise, so that the improvement is small.
\end {abstract}

%\pacs {98.80.Cq, 98.80.Es, 98.65.Dx, 98.70.Vc}


\section {Introduction} \label {sec:intro}
The quest to measure primordial non-Gaussianity (PNG) has been a thriving field for the past decade. 
PNG has long been considered an open window onto the physics of the early universe, affording the exciting possibility of ruling out the canonical slow-roll inflation model and finding evidence for new primordial physics \citep{Byrnes:2010a}.

PNG has been traditionally probed with the bispectrum of the CMB anisotropies, which is expected to vanish at first order in a fully Gaussian universe. While possible hints of departures from Gaussianity have occasionally appeared at low significance from analyses based on WMAP data \citep{2008PhRvL.100r1301Y}, PNG remained weakly constrained \citep{2012arXiv1212.5225B,2012arXiv1212.5226H}, encouraging a sustained growth of theoretical models of inflation producing non-Gaussian initial conditions \citep{2010JCAP...12..030S}. The constraints on PNG from the CMB bispectrum have now improved dramatically with the first-year release of the Planck CMB data, from whose bispectrum it was found that local $\fnl = 2.7 \pm 5.8$ (68 \% c.l.) \citep{2013arXiv1303.5084P} after subtraction of non-primordial contributions; this result has put significant pressure onto multi-field inflation, reducing the scope of possible discoveries. However, some non-Gaussianity at the level of $\fnl \sim 1$ is expected even in the canonical model \citep{2004PhR...402..103B}, meaning it is worthwhile to look for methods to further improve the existing constraints. Furthermore, it is worth cross-checking all constraints with independent methods.

The discovery by Dalal and collaborators \citep{Dalal:2008a, Matarrese:2008a, Slosar:2008a,2009MNRAS.396...85D,2010CQGra..27l4011D,2010AdAst2010E..89D,2010PhRvD..82j3529D,Giannantonio:2010a,Valageas:2010a,Desjacques:2011a,2011PhRvD..84f3512D} that the bias of dark matter haloes and galaxies becomes strongly scale-dependent in the presence of PNG opened up a new avenue for PNG measurement. Constraints on PNG from bias measurements of different galaxy samples were found to be competitive, and comparable with, CMB bispectrum results before the Planck data release \citep{Slosar:2008a,Xia:2010a,Xia:2010b,Xia:2011a}. The strongest robust constraints obtained with this technique were recently described by \citet{2013arXiv1303.1349G}, where a compilation of six galaxy catalogues and their external correlations with the CMB temperature anisotropies was used to measure $\fnl = 5 \pm 21 $ (68 \% c.l.) for the local configuration under the most conservative assumptions.

The measurement of PNG from scale-dependent large-scale bias is complicated by observational systematics, such as stellar contamination in galaxy samples, which acts to produce large-scale power mimicking a PNG signal \citep{Ross:2011a,Ross12}. If we can model the phase information of the systematic we can ignore the affected modes \citep{2013MNRAS.435.1857L}, or we can weight the galaxies to create an unbiased field \citep{Ross13fnl}. However, these systematics can be most easily controlled by using measurements of cross-correlations between different galaxy samples or between samples and other data that trace the density field, for which we expect uncorrelated observational systematics.

In this letter we focus on a newer addition to existing large-scale structure (LSS) methods to measure PNG: the galaxy bias and thus PNG can also be measured by cross-correlating galaxies and the matter density field, reconstructed from gravitational lensing \citep{2009PhRvD..80l3527J}. The special case of CMB lensing \citep{2006PhR...429....1L} is particularly useful, because it allows consistent tomographic correlations with galaxy surveys; early forecasts showed that this method can provide competitive PNG constraints \citep{2009PhRvD..80l3527J,2010PhRvD..82b3517T, 2012PhRvD..85d3518T}. CMB lensing maps have now been reconstructed, and their correlations with galaxy surveys have been confirmed, using data from the Planck satellite \citep{2013arXiv1303.5077P}, the South Pole Telescope \citep{2012ApJ...756..142V, 2012ApJ...753L...9B}, and the Atacama Cosmology Telescope \citep{2011PhRvL.107b1301D, 2012PhRvD..86h3006S}. We can now for the first time apply this method to constrain PNG using public CMB lensing data from Planck.

We update the existing analysis by \citet{2013arXiv1303.1349G} as follows: in addition to the density-density correlations between six galaxy catalogues, and to their cross-correlations with the CMB temperature anisotropies that we update to Planck, we also measure and use their cross-correlations with the recently released Planck CMB lensing map \citep{2013arXiv1303.5077P}. We test how these correlations can improve the combined PNG constraints, and we show that they also represent an additional, partially independent cross-check on the results.

\section {Theory}
In this letter we only consider the simplest local PNG model, parametrized as 
\ba \label{eq:fnl101}
\Phi (\mathbf{x}, z_{\star}) &=& \varphi(\mathbf{x}, z_{\star}) + \fnl \, \left[ \varphi^2(\mathbf{x}, z_{\star}) - \langle \varphi^2 \rangle(z_{\star}) \right] 
\ea
where $\fnl$ 
quantifies the amount of PNG. Here $\Phi(\mathbf{x},z_{\star})$ is the potential at primordial times $z_{\star}$ and $\varphi$ an auxiliary Gaussian potential. In the presence of PNG, the bias becomes scale-dependent, and is well described by a correction
\be
\Delta b^{\mathrm{loc}} (k, \fnl) = 2 \, \delta_c \, b_L \, \fnl / \alpha(k) \, ,
\ee
where $\alpha(k,z) = \frac{2 \, k^2 \, T(k) \, D(z)}{3 \, \Omega_m \, H_0^2} \, \frac{g(0)}{g(z_{\star})}$, $T(k)$ is the density transfer function, $D(z)$ is the linear growth function, $g(z) \propto (1+z) \, D(z)$ is the potential growth function,    $\delta_c = 1.686$ is the spherical collapse threshold and $b_L \equiv b_1 - 1 $ is the Lagrangian bias. 

We constrain $\fnl$ via the galaxy bias, as measured by the 2D angular correlation functions $w^{AB}(\vartheta)$ between all pairs of maps $A,B$, whose theoretical predictions are calculated numerically as a Legendre transformation of the corresponding angular power spectra $C_l^{AB} = (2/\pi) \, \int\, dk \,  k^2 \, P(k) \, W_l^A(k) \, W_l^B(k)$. The sources $W_l(k)$ describe the redshift projection over the survey visibility function $dN/dz(z)$ of the physical sources, which are different for galaxy counts, ISW, and CMB lensing, calculated using a modified version of \textsc{Camb} \citep {Lewis:2000a}.

\begin{figure}
\begin{center}
\includegraphics[width=\linewidth, angle=0]{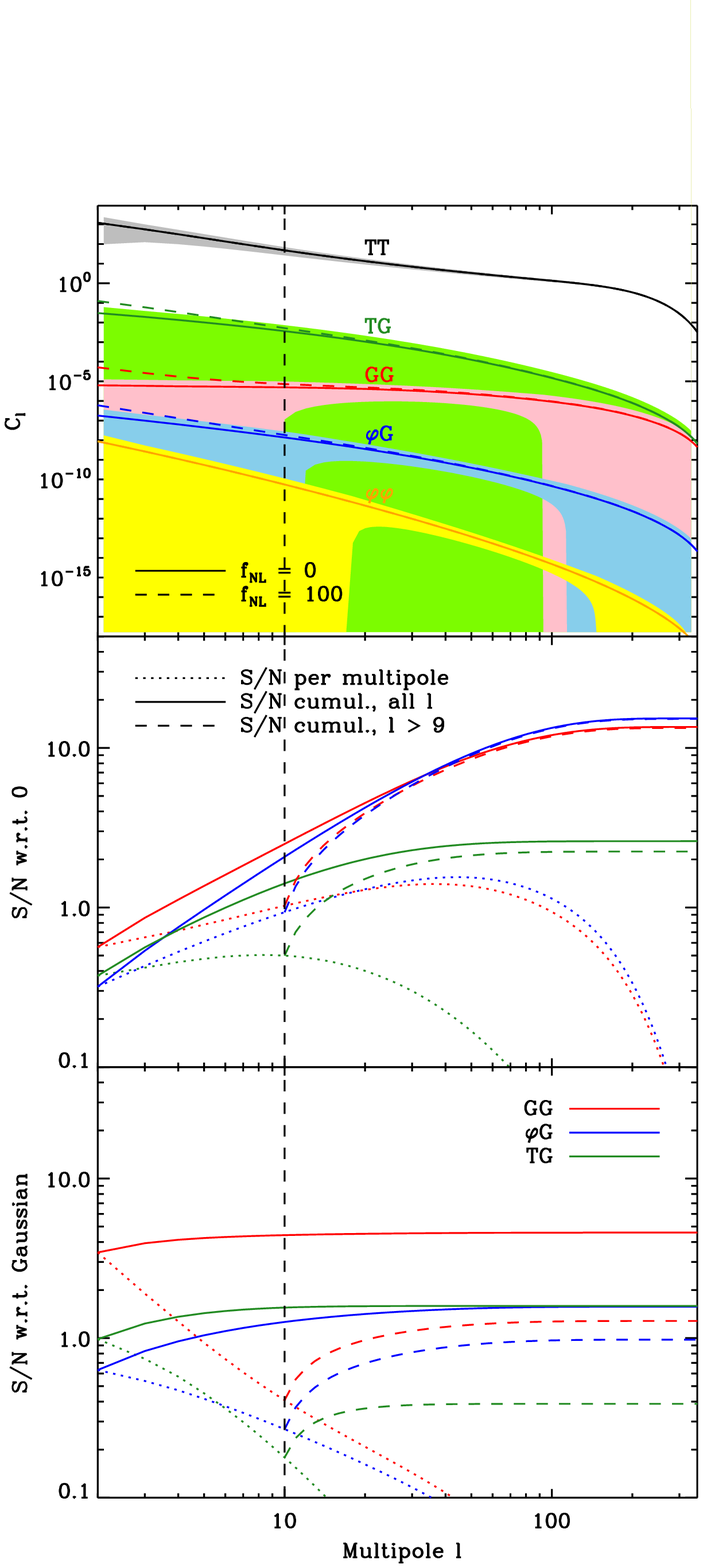}
\caption{Linear theory predictions of the angular power spectra we consider including their theoretical uncertainties given by cosmic variance, shot noise, and intrinsic lensing noise (top panel) for two models with $\fnl = 0, 100$, and signal-to-noise estimations for a single galaxy catalogue. We assume the specifications of the NVSS survey and the WMAP7 best-fit cosmology, and apply a Healpix smoothing for $N_{\mathrm{side}} = 64$ ($\sim 50$ arcmin) as used in the analysis. The central panel shows the absolute signal-to-noise ratio of the three spectra we measure, per multipole and cumulative (the latter starting from $l = 2$ or from $l = 10$). The bottom panel shows the approximated detection power of a model with $\fnl = 100$, defined as the signal-to-noise ratio of the difference between the two models.}
\label{fig:sn}
\end{center}
\end{figure}

The total signal-to-noise ratio to be expected for a single galaxy catalogue and its external correlations (GG, TG, $\varphi G$) is shown in Fig.~\ref{fig:sn} for one case (corresponding to NVSS specifications; see below for details of the data sets we consider). Here we include in the theoretical uncertainties cosmic variance, shot noise, and the intrinsic lensing noise provided by Planck. We can see in the central panel that the total signal-to-noise of both TG and $\varphi G$ signals is barely affected by the modes at $l < 10$; the constraining power on $\fnl$ however, defined as the signal-to-noise ratio of the difference between a Gaussian and a non-Gaussian model, is reduced if the largest scales are excluded, as the scale-dependent bias is most visible precisely for these modes (bottom panel). We can also see that the constraining power on $\fnl$ of the new $\varphi G$ correlations should be comparable with the TG part if all modes were available, and marginally less if using the cut data at $l > 9 $ only. The galaxy auto-correlation functions (ACFs) are expected to constrain $\fnl$ more strongly because the bias enters in quadrature in this case. It is finally important to notice from the central panel that the total signal-to-noise of the $\varphi G$ correlations is actually high, comparable with the ACF; but the largest contribution arise at smaller scales, thus limiting the constraining power on the scale-dependent bias.

\section {Data}
We consider the compilation of six galaxy catalogues introduced by \citet{Giannantonio:2008a}, updated in \citet{2012MNRAS.426.2581G} and used to constrain PNG using density and density-CMB temperature correlations in \citet{2013arXiv1303.1349G}. Briefly, this consists of the IR galaxies of 2MASS at a median redshift $z \simeq 0.1$, the radio-galaxies of NVSS and X-ray background of HEAO (both spanning a broad redshift range), and three photometric samples from the Sloan Digital Sky Survey (SDSS), i.e. the main galaxies at $z \simeq 0.3$, the luminous red galaxies (LRGs) from the photometric CMASS sample from Data Release 8 (DR8) at $z \simeq 0.5$ and the DR6 photometric quasars, which also feature a broad redshift distribution. 

We replace the previously used WMAP CMB data with the newly released Planck maps. We use the temperature \textsc{Smica} map with the  strictest provided galaxy mask, as well as the CMB lensing map reconstructed from the off-diagonal covariances between different multipoles in the temperature map together with its mask. The Planck collaboration removed information for the largest scales (modes with $l < 10$) from this map: 
although the scale-dependent bias affects mostly the largest scales, as shown in Fig. ~\ref{fig:sn}, the increased noise means that we do not expect a drastic degradation in constraining power on $\fnl$. We test this further with mock data below.

We first measure all projected two-point angular correlation functions $w^{g_i g_j}(\vartheta)$ between pairs of  catalogues $i,j$ at angular separations $0 \le \vartheta \le 12 $ deg using a pixel-based estimator within the \textsc{Healpix} scheme \citep{2005ApJ...622..759G} at $N_{\mathrm{side}} = 64$ (pixel size $\sim 50$ arcmin): this yields 21 correlation functions. Some of the auto-correlation functions (ACFs) present an excess power at large angular separations compared with the Gaussian \LCDM~predictions, especially the quasars and the NVSS galaxies; a detailed analysis of the systematics of these samples was presented in \citet{2013arXiv1303.1349G}, where it was shown that such signals are likely due to residual systematic contaminations, as also demonstrated by \citet{Pullen_qso,2013MNRAS.435.1857L}. Following these systematics tests, it was decided to take the most conservative approach and to keep the raw NVSS data uncorrected for the existing r.a. and declination-dependent systematics, to avoid the risk of biasing the constraints on $\fnl$. The NVSS and QSO ACFs are then discarded from the cosmological analysis.
We adopt the same choice here, while keeping all the cross-correlations between the different data sets.

\begin{figure}
\begin{center}
\includegraphics[width=\linewidth, angle=0]{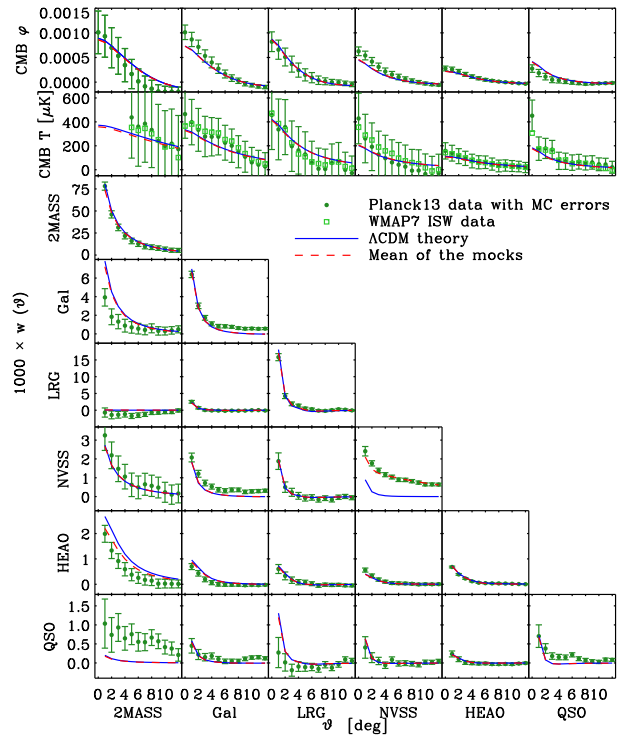}
\caption{The full extended data set used in this analysis. The first row shows the new set of galaxy-CMB lensing correlation functions. The second row is the ISW effect (compared between WMAP and Planck), and the remaining rows are the galaxy-galaxy correlation functions. Error bars are Monte Carlos and are highly correlated. The ISW 2MASS error bars are $0.5\sigma$. The ACF of the raw NVSS data presents a significant excess power with respect to the \LCDM~expectations, which is modeled by adding to the mocks the r.a. and dec density fluctuations observed in the data. The NVSS and QSO ACFs are not used for the cosmological results due to their known systematics.}
\label{fig:data}
\end{center}
\end{figure}

We then measure the six cross-correlation functions (CCFs) between the galaxy catalogues and the CMB temperature anisotropies $w^{T g_i} (\vartheta)$, updating our analysis to the Planck first year data release \citep{2013arXiv1303.5062P}. The measured level of these correlations is consistent with the assumption that they are produced by the integrated Sachs-Wolfe effect (ISW). This corresponds to the `fair' sample of \citet{2013arXiv1303.1349G}.
We finally add to our data set the six CCFs between the galaxy catalogues and the Planck CMB lensing map \citep{2013arXiv1303.5077P}: $w^{\varphi g_i} (\vartheta)$. These correlations allow a redshift tomography of the CMB lensing sources, effectively mapping the dark matter distribution in redshift bins. 
We thus obtain the 33 correlation functions shown in Fig.~\ref{fig:data}.
Notice that we have nulled the angular power spectrum at $l<10$ in the galaxy-lensing spectrum for consistency with the Planck data. We calculate the covariance matrix between all $33 \times 13 = 429$ data points using a Monte Carlo method, generating 10,000 realisations based on a fiducial Gaussian \LCDM~model, including shot noise in the counts, the intrinsic lensing noise from Planck, and all expected correlations between the maps \citep[see Appendix of][]{Giannantonio:2008a}. 
We also include the r.a. and declination-dependent systematics in the mock NVSS data, so that the mean of the mocks used to estimate the covariance agrees with the observed ACF \citep{2013arXiv1303.1349G}.

\section {Results} We  calculate the likelihood of the theoretical parameters given the first-year Planck temperature power spectrum (with WMAP polarisation) to impose tight priors on most cosmological parameters, while our compilation of correlation functions will constrain $\fnl$.
We consider different subsets of our data, exploring the parameter space with a modified version of the latest \textsc{Cosmomc} code \citep{Lewis:2002a}, including the official Planck likelihood code. As discussed in more detail in \citet{2013arXiv1303.1349G}, in addition to the standard \LCDM~cosmological parameters, we always vary a set of ten nuisance parameters to account for uncertainties in our modelling of the data: one free bias parameter for each $i$-th catalogue $b_0^i$, one stellar contamination fraction $\kappa^i$ for each of the SDSS samples, and one PSF smoothing for the HEAO data $\alpha_{\mathrm{HEAO}}$. As in \citet{2013arXiv1303.1349G} we assume that the Gaussian part of the bias of most samples evolves as $ b_1^i(z) = 1 + \left[ b_0^i - 1 \right] / {D^2(z)} $, while for the quasars we assume $b_1^{\mathrm{QSO}}(z) =  b_0^{\mathrm{QSO}} / {D^{1.6}(z)}$; for further details see \citet{2013arXiv1303.1349G}, where it was found that the results do not depend too strongly on these assumptions.
We also use the standard set of nuisance parameters introduced in the Planck likelihood package.

We summarise our results in Table~\ref{tab:results}. When using the Planck TT data with WMAP polarisation (WP), and the GG correlation functions only, we find $\fnl = 15 \pm 29$ (all results at $1 \sigma$). The addition of the LSS-CMB temperature correlations (ISW) improves this to $\fnl = 14 \pm 25$. Note that this error is consistent with, although slightly worse than the error found in \citet{2013arXiv1303.1349G}: a consequence of the different corrections assumed for the CMB data. As we are considering large-scales only, the WMAP and Planck data  provide similar signal-to-noise.
If instead of the ISW we add the CMB lensing correlations, we find $\fnl = 11 \pm 23$, while the final, fully combined results (including all correlations) yields $\fnl = 12 \pm 21$. 

To better compare the constraining power of the different parts of our data set, we also test the results on $\fnl$ when using the GG, TG, and $\varphi G$ parts only. In order to make the comparison more meaningful for these runs, we included Gaussian priors on the bias and stellar contamination parameters equal to their posteriors from the full run.  The results presented in Table~\ref{tab:results} show that the constraining power on $\fnl$ of the TG part is marginally stronger than the $\varphi G$ at $l > 9$, while the GG part is a factor of $\sim 3$ better. This is in qualitative agreement with our signal-to-noise calculations shown in Fig.~\ref{fig:sn}.
 
\begin{table}
\begin{center}
\begin{tabular}{ccr}
% \hline 
\mbox{Data: Planck TT, WP, and\ }  & \ Priors\  & \ $\fnl$ (68\%) \\
\hline
\\
 $GG$           & $b_0^i , \kappa_i$ & $ 12 \pm 23 $ \\
 $TG$           & $b_0^i , \kappa_i$ & $ 46 \pm 68 $ \\
 $\varphi G$ & $b_0^i , \kappa_i$ & $ 12 \pm 71 $ \\
%\hline
Mock all-$l$ $\varphi G$ & $b_0^i , \kappa_i$ & $ \pm 53 $ \\
\\
 $GG$                         & none & $ 15 \pm 29 $ \\
 $GG+TG$                  & none & $ 14 \pm 25 $ \\
 $GG+\varphi G$        & none & $ 11 \pm 23 $ \\
 $GG+\varphi G+TG$,& none & $  12 \pm 21 $ \\
Mock all-$l$, $GG+\varphi G+TG$& none & $ \pm 19 $ \\
as above, no intrinsic noise & none & $ \pm 14 $ \\
\hline
\end{tabular}
\caption{Measurements of $\fnl$, for different combinations of data. Where the data come from mock catalogues including lensing on scales $l<10$, only the errors are provided.}
\label{tab:results}
\end{center}
\end{table}

\section {Forecast for uncut lensing maps}
\begin{figure}
\begin{center}
\includegraphics[width=\linewidth, angle=0]{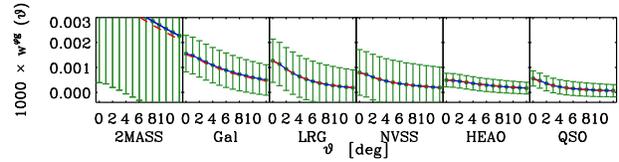}
\caption{The mock $\varphi G$ correlations, for the case of uncut CMB lensing data. We set the mock data to be equal to the fiducial \LCDM~model, and the covariance matrix has been re-calculated using the full angular power spectrum at all $l$.}
\label{fig:data_allell}
\end{center}
\end{figure}

As previously mentioned, the publicly available CMB lensing map reconstructed by the Planck collaboration has had modes $l < 10$ removed. As it is known that in the presence of PNG scale-dependent bias is strongest on the largest scales, here we address the question of how much better would our constraints be if we could use the full uncut CMB lensing data. For this purpose, we replace the measured $w^{\varphi g_i} (\vartheta)$ data points with mock data that we set equal to our fiducial \LCDM~model. We also generate a new covariance matrix, where the input fiducial model does include all multipoles in the LSS-CMB lensing correlations. We show the modified data set in Fig.~\ref{fig:data_allell}, where we can see that both the signal and the error bars in the CMB lensing correlations have significantly increased.

This can be readily understood by remembering that, in the simplified case of cosmic variance-dominated errors, the variance is proportional to the angular power spectrum, which steeply increases at the smallest multipoles in the CMB lensing case. In addition to this, the intrinsic lensing noise of Planck is also large compared with the signal on these scales \citep[see Fig.~1 in][]{{2013arXiv1303.5077P}}. Thus, when projecting to real space, the inclusion of the modes at $2 < l < 10$ will bring a large contribution for both signal and noise, as shown in Fig.~\ref{fig:sn}. We have run the full likelihood analysis on this modified data set, and find marginally improved results, with error on $\fnl$, $\pm 19$. This is again in agreement with the signal-to-noise projection of Fig.~\ref{fig:sn}. We finally test how much would the results improve if we had an ideal experiment without any intrinsic lensing noise: in this case we find an error $\pm 14$ using all of the data.

\section {Conclusions} \label{sec:conclusion}
We have applied a new method to improve the large-scale structure constraints on primordial non-Gaussianity, using the cross-correlations of galaxy catalogues with CMB lensing maps. New maps from the Planck satellite were used to measure the PNG parameter $\fnl$, finding similar errors to those from ISW based bias measurements. Consequently, the final combined measurements of $\fnl$ are only marginally improved by including density-CMB lensing correlations in addition to density and density-CMB temperature correlations. We have investigated the penalising effects of cosmic variance, intrinsic lensing noise, and cuts imposed on the Planck CMB lensing maps, finding consistency between results and expectations. Combining all of our measurements, we find $\fnl = 12 \pm 21$ ($1\sigma$).

The addition of the CMB lensing correlations provides an important consistency check for $\fnl$ measurements, as it is expected to be affected by different systematics than ISW, galaxy-galaxy correlation, and bispectrum based measurements. The method presented in this letter serves also as a preliminary exercise for the Dark Energy Survey (DES; {\tt www.darkenergysurvey.com}), to which we will apply a similar analysis in the near future. Beyond PNG, the consistent combination of internal and external correlation functions of the LSS represents a powerful way to extract the most cosmological information, and to reconstruct the evolution of the Universe at the perturbative level. Based on our analysis, the addition of CMB lensing is expected to provide more powerful cosmological measurements on smaller scales than those used here to contain the PNG signal. Thus, future analyses of Dark Energy and of neutrino masses will be particularly interesting \citep{2013arXiv1311.0905P}. As clustering, the ISW, and gravitational lensing are sensitive to different combinations of the gravitational potentials and their derivatives, their combination could also provide a powerful tool to constrain the history of gravity and structure formation.

\section*{Acknowledgement}
We thank Aur\'{e}lien Beno\^{i}t-Levy and Pablo Fosalba for useful discussions on the Planck CMB lensing data. We also thank Eiichiro Komatsu and Bj\"orn S\"orgel for useful comments. TG acknowledges the Rechenzentrum Garching of the Max Planck Society for computational resources.
 WJP acknowledges support from the UK Science \& Technology Facilities Council (STFC) through the consolidated grant ST/K0090X/1, and from the European Research Council through the ``Starting Independent Research'' grant 202686, MDEPUGS.

\appendix

\bibliographystyle{mn2e}
\bibliography{ms}

\begin{thebibliography}{43}
\expandafter\ifx\csname natexlab\endcsname\relax\def\natexlab#1{#1}\fi

\bibitem[{{Bartolo} {et~al}\mbox{.}(2004){Bartolo}, {Komatsu}, {Matarrese}, \&
  {Riotto}}]{2004PhR...402..103B}
{Bartolo} N., {Komatsu} E., {Matarrese} S., {Riotto} A., 2004, \physrep, 402,
  103

\bibitem[{{Bennett} {et~al}\mbox{.}(2013){Bennett}
  {et~al.}}]{2012arXiv1212.5225B}
{Bennett} C.~L., {et~al.}, 2013, \apjs, 208, 20

\bibitem[{{Bleem} {et~al}\mbox{.}(2012){Bleem} {et~al.}}]{2012ApJ...753L...9B}
{Bleem} L.~E., {et~al.}, 2012, \apjl, 753, L9

\bibitem[{{Byrnes} \& {Choi}(2010)}]{Byrnes:2010a}
{Byrnes} C.~T., {Choi} K.-Y., 2010, Advances in Astronomy, 2010

\bibitem[{{Dalal} {et~al}\mbox{.}(2008){Dalal}, {Dor{\'e}}, {Huterer}, \&
  {Shirokov}}]{Dalal:2008a}
{Dalal} N., {Dor{\'e}} O., {Huterer} D., {Shirokov} A., 2008, \prd, 77, 123514

\bibitem[{{Das} {et~al}\mbox{.}(2011){Das} {et~al.}}]{2011PhRvL.107b1301D}
{Das} S., {et~al.}, 2011, Physical Review Letters, 107, 021301

\bibitem[{{Desjacques} {et~al}\mbox{.}(2010){Desjacques}, {Crocce},
  {Scoccimarro}, \& {Sheth}}]{2010PhRvD..82j3529D}
{Desjacques} V., {Crocce} M., {Scoccimarro} R., {Sheth} R.~K., 2010, \prd, 82,
  103529

\bibitem[{{Desjacques} {et~al}\mbox{.}(2011{\natexlab{a}}){Desjacques},
  {Jeong}, \& {Schmidt}}]{Desjacques:2011a}
{Desjacques} V., {Jeong} D., {Schmidt} F., 2011{\natexlab{a}}, \prd, 84, 061301

\bibitem[{{Desjacques} {et~al}\mbox{.}(2011{\natexlab{b}}){Desjacques},
  {Jeong}, \& {Schmidt}}]{2011PhRvD..84f3512D}
{Desjacques} V., {Jeong} D., {Schmidt} F., 2011{\natexlab{b}}, \prd, 84, 063512

\bibitem[{{Desjacques} \& {Seljak}(2010{\natexlab{a}})}]{2010CQGra..27l4011D}
{Desjacques} V., {Seljak} U., 2010{\natexlab{a}}, Classical and Quantum
  Gravity, 27, 124011

\bibitem[{{Desjacques} \& {Seljak}(2010{\natexlab{b}})}]{2010AdAst2010E..89D}
{Desjacques} V., {Seljak} U., 2010{\natexlab{b}}, Advances in Astronomy, 2010

\bibitem[{{Desjacques} {et~al}\mbox{.}(2009){Desjacques}, {Seljak}, \&
  {Iliev}}]{2009MNRAS.396...85D}
{Desjacques} V., {Seljak} U., {Iliev} I.~T., 2009, \mnras, 396, 85

\bibitem[{{Giannantonio} {et~al}\mbox{.}(2012){Giannantonio}, {Crittenden},
  {Nichol}, \& {Ross}}]{2012MNRAS.426.2581G}
{Giannantonio} T., {Crittenden} R., {Nichol} R., {Ross} A.~J., 2012, \mnras,
  426, 2581

\bibitem[{{Giannantonio} \& {Porciani}(2010)}]{Giannantonio:2010a}
{Giannantonio} T., {Porciani} C., 2010, \prd, 81, 063530

\bibitem[{{Giannantonio} {et~al}\mbox{.}(2013){Giannantonio}, {Ross},
  {Percival}, {Crittenden}, {Bacher}, {Kilbinger}, {Nichol}, \&
  {Weller}}]{2013arXiv1303.1349G}
{Giannantonio} T., {Ross} A.~J., {Percival} W.~J., {Crittenden} R., {Bacher}
  D., {Kilbinger} M., {Nichol} R., {Weller} J., 2013, Phys. Rev. D, in press,
  arXiv:1303.1349

\bibitem[{{Giannantonio} {et~al}\mbox{.}(2008){Giannantonio}, {Scranton},
  {Crittenden}, {Nichol}, {Boughn}, {Myers}, \&
  {Richards}}]{Giannantonio:2008a}
{Giannantonio} T., {Scranton} R., {Crittenden} R.~G., {Nichol} R.~C., {Boughn}
  S.~P., {Myers} A.~D., {Richards} G.~T., 2008, \prd, 77, 123520

\bibitem[{{G{\'o}rski} {et~al}\mbox{.}(2005){G{\'o}rski}, {Hivon}, {Banday},
  {Wandelt}, {Hansen}, {Reinecke}, \& {Bartelmann}}]{2005ApJ...622..759G}
{G{\'o}rski} K.~M., {Hivon} E., {Banday} A.~J., {Wandelt} B.~D., {Hansen}
  F.~K., {Reinecke} M., {Bartelmann} M., 2005, \apj, 622, 759

\bibitem[{{Hinshaw} {et~al}\mbox{.}(2013){Hinshaw}
  {et~al.}}]{2012arXiv1212.5226H}
{Hinshaw} G., {et~al.}, 2013, \apjs, 208, 19

\bibitem[{{Jeong} {et~al}\mbox{.}(2009){Jeong}, {Komatsu}, \&
  {Jain}}]{2009PhRvD..80l3527J}
{Jeong} D., {Komatsu} E., {Jain} B., 2009, \prd, 80, 123527

\bibitem[{{Leistedt} {et~al}\mbox{.}(2013){Leistedt}, {Peiris}, {Mortlock},
  {Benoit-L{\'e}vy}, \& {Pontzen}}]{2013MNRAS.435.1857L}
{Leistedt} B., {Peiris} H.~V., {Mortlock} D.~J., {Benoit-L{\'e}vy} A.,
  {Pontzen} A., 2013, \mnras, 435, 1857

\bibitem[{{Lewis} \& {Bridle}(2002)}]{Lewis:2002a}
{Lewis} A., {Bridle} S., 2002, \prd, 66, 103511

\bibitem[{{Lewis} \& {Challinor}(2006)}]{2006PhR...429....1L}
{Lewis} A., {Challinor} A., 2006, \physrep, 429, 1

\bibitem[{{Lewis} {et~al}\mbox{.}(2000){Lewis}, {Challinor}, \&
  {Lasenby}}]{Lewis:2000a}
{Lewis} A., {Challinor} A., {Lasenby} A., 2000, \apj, 538, 473

\bibitem[{{Matarrese} \& {Verde}(2008)}]{Matarrese:2008a}
{Matarrese} S., {Verde} L., 2008, \apjl, 677, L77

\bibitem[{{Pearson} \& {Zahn}(2013)}]{2013arXiv1311.0905P}
{Pearson} R., {Zahn} O., 2013, ArXiv e-prints, arxiv:1311.0905

\bibitem[{{Planck Collaboration}(2013{\natexlab{a}})}]{2013arXiv1303.5062P}
{Planck Collaboration}, 2013{\natexlab{a}}, ArXiv e-prints, arxiv:1303.5062

\bibitem[{{Planck Collaboration}(2013{\natexlab{b}})}]{2013arXiv1303.5077P}
{Planck Collaboration}, 2013{\natexlab{b}}, ArXiv e-prints, arxiv:1303.5077

\bibitem[{{Planck Collaboration}(2013{\natexlab{c}})}]{2013arXiv1303.5084P}
{Planck Collaboration}, 2013{\natexlab{c}}, ArXiv e-prints, arxiv:1303.5084

\bibitem[{{Pullen} \& {Hirata}(2013)}]{Pullen_qso}
{Pullen} A.~R., {Hirata} C.~M., 2013, \pasp, 125, 705

\bibitem[{{Ross} {et~al}\mbox{.}(2011){Ross} {et~al.}}]{Ross:2011a}
{Ross} A.~J., {et~al.}, 2011, \mnras, 417, 1350

\bibitem[{{Ross} {et~al}\mbox{.}(2012){Ross} {et~al.}}]{Ross12}
{Ross} A.~J., {et~al.}, 2012, \mnras, 424, 564

\bibitem[{{Ross} {et~al}\mbox{.}(2013){Ross} {et~al.}}]{Ross13fnl}
{Ross} A.~J., {et~al.}, 2013, \mnras, 428, 1116

\bibitem[{{Sherwin} {et~al}\mbox{.}(2012){Sherwin}
  {et~al.}}]{2012PhRvD..86h3006S}
{Sherwin} B.~D., {et~al.}, 2012, \prd, 86, 083006

\bibitem[{{Slosar} {et~al}\mbox{.}(2008){Slosar}, {Hirata}, {Seljak}, {Ho}, \&
  {Padmanabhan}}]{Slosar:2008a}
{Slosar} A., {Hirata} C., {Seljak} U., {Ho} S., {Padmanabhan} N., 2008, \jcap,
  8, 31

\bibitem[{{Suyama} {et~al}\mbox{.}(2010){Suyama}, {Takahashi}, {Yamaguchi}, \&
  {Yokoyama}}]{2010JCAP...12..030S}
{Suyama} T., {Takahashi} T., {Yamaguchi} M., {Yokoyama} S., 2010, \jcap, 12, 30

\bibitem[{{Takeuchi} {et~al}\mbox{.}(2010){Takeuchi}, {Ichiki}, \&
  {Matsubara}}]{2010PhRvD..82b3517T}
{Takeuchi} Y., {Ichiki} K., {Matsubara} T., 2010, \prd, 82, 023517

\bibitem[{{Takeuchi} {et~al}\mbox{.}(2012){Takeuchi}, {Ichiki}, \&
  {Matsubara}}]{2012PhRvD..85d3518T}
{Takeuchi} Y., {Ichiki} K., {Matsubara} T., 2012, \prd, 85, 043518

\bibitem[{{Valageas}(2010)}]{Valageas:2010a}
{Valageas} P., 2010, \aap, 514, A46

\bibitem[{{van Engelen} {et~al}\mbox{.}(2012){van Engelen}
  {et~al.}}]{2012ApJ...756..142V}
{van Engelen} A., {et~al.}, 2012, \apj, 756, 142

\bibitem[{{Xia} {et~al}\mbox{.}(2011){Xia}, {Baccigalupi}, {Matarrese},
  {Verde}, \& {Viel}}]{Xia:2011a}
{Xia} J.-Q., {Baccigalupi} C., {Matarrese} S., {Verde} L., {Viel} M., 2011,
  \jcap, 8, 33

\bibitem[{{Xia} {et~al}\mbox{.}(2010{\natexlab{a}}){Xia}, {Bonaldi},
  {Baccigalupi}, {De Zotti}, {Matarrese}, {Verde}, \& {Viel}}]{Xia:2010a}
{Xia} J.-Q., {Bonaldi} A., {Baccigalupi} C., {De Zotti} G., {Matarrese} S.,
  {Verde} L., {Viel} M., 2010{\natexlab{a}}, \jcap, 8, 13

\bibitem[{{Xia} {et~al}\mbox{.}(2010{\natexlab{b}}){Xia}, {Viel},
  {Baccigalupi}, {De Zotti}, {Matarrese}, \& {Verde}}]{Xia:2010b}
{Xia} J.-Q., {Viel} M., {Baccigalupi} C., {De Zotti} G., {Matarrese} S.,
  {Verde} L., 2010{\natexlab{b}}, \apjl, 717, L17

\bibitem[{{Yadav} \& {Wandelt}(2008)}]{2008PhRvL.100r1301Y}
{Yadav} A.~P.~S., {Wandelt} B.~D., 2008, Physical Review Letters, 100, 181301

\end{thebibliography}

\label{lastpage}

\end{document}